\documentclass[
    reprint,
    superscriptaddress,
    amsmath,
    amssymb,
    aps,
	prl]{revtex4-2}

\usepackage{graphicx}
\usepackage{dcolumn}
\usepackage{bm}
\usepackage{hyperref}
\usepackage{siunitx}
\usepackage{blindtext}
\usepackage{comment}

\DeclareSIUnit{\rad}{rad}
\DeclareSIUnit{\dBm}{dBm}

\usepackage{cleveref}
\crefname{section}{Sec.}{Secs.}
\Crefname{section}{Section}{Sections}
\crefname{equation}{Eq.}{Eqs.}
\crefname{figure}{Fig.}{Figs.}
\Crefname{figure}{Figure}{Figures}

\graphicspath{{images}}

\begin{document}

\preprint{Version 2026.09.07}

\renewcommand{\d}{\ensuremath{\mathrm{d}}}

\newcommand{\stbg}[1]{{\color[rgb]{1,0,0} #1}}
\newcommand{\rs}[1]{{\color[rgb]{0,0.7,0.7} #1}}
\newcommand{\fh}[1]{{\color[rgb]{0,0,1} #1}}
\newcommand{\move}[1]{{\color[rgb]{0,0,1} #1}}
\newcommand{\remove}[1]{{\color[rgb]{0,0.5,0} #1}}

\title{\textbf{Quantifying thermal and driven magnon populations with femtosecond noise correlation spectroscopy}}

\author{F.~S.~Herbst}
    \affiliation{Department of Physics, University of Konstanz, 78457 Konstanz, Germany}
\author{M.~A.~Weiss}
    \affiliation{Department of Physics, University of Konstanz, 78457 Konstanz, Germany}
\author{A.~Leitenstorfer}
    \affiliation{Department of Physics, University of Konstanz, 78457 Konstanz, Germany}
\author{M.~Lammel}
    \affiliation{Department of Physics, University of Konstanz, 78457 Konstanz, Germany}
\author{N.~Beaulieu}
    \affiliation{LabSTICC, CNRS, Université de Bretagne Occidentale, Brest 29238, France}
\author{J.~Ben~Youssef}
    \affiliation{LabSTICC, CNRS, Université de Bretagne Occidentale, Brest 29238, France}
\author{R.~Schlitz}
    \affiliation{Department of Physics, University of Konstanz, 78457 Konstanz, Germany}
\author{S.~T.~B.~Goennenwein}
    \affiliation{Department of Physics, University of Konstanz, 78457 Konstanz, Germany}

\date{\today}

\begin{abstract}

Precise knowledge of the total number of magnons, including both coherent and incoherent (e.g. thermal) excitations, is imperative for the advancement of fundamental spin-wave physics and the development of next-generation magnonic devices.
In particular, quantifying magnons is key to understanding magnon transport phenomena, the nonlinear regime, or ultrafast magnetization dynamics.
Typically, incoherent magnons are accessed by frequency-domain techniques, which lack the temporal resolution required for ultrafast processes, while ultrafast time-domain methods are generally sensitive only to the coherent dynamics.
In this work, we demonstrate that femtosecond noise correlation spectroscopy enables a fully quantitative, time-domain measurement of both thermal and coherently excited magnon modes in bismuth-substituted yttrium iron garnet driven by a free-running microwave.
We model the experimental data and extract the magnon number by simulating the magnon band structure of the sample, the magneto-optical response function, and the optical spot size used in the experiment. 
Our analysis establishes a connection between magnon mode calculations and experimentally accessible magnetic properties and fiducially reproduces the waveform and amplitude of the magneto-optical correlation signal for different experimental conditions. 
These results open a new pathway towards the optical tomography of magnon modes in non-linear or non-equilibrium conditions and can be readily extended to study ultrafast incoherent dynamics in other condensed matter systems. 

\end{abstract}

\maketitle

Magnetic systems exhibit a rich variety of dynamic phenomena on ultrafast timescales, which are of both fundamental and technological interest \cite{kirilyukUltrafastOpticalManipulation2010}.
Spintronic devices that leverage these dynamics prospectively offer high-speed and low-power alternatives to conventional electronics \cite{chumakMagnonSpintronics2015, manchonCurrentinducedSpinorbitTorques2019, krizakovaSpinorbitTorqueSwitching2022}.
In this context, the quanta of spin waves -- magnons -- are discussed as promising candidates for information carriers 
\cite{chumakMagnonSpintronics2015, chumakAdvancesMagneticsRoadmap2022,flebus2024MagnonicsRoadmap2024, demidovSpinOrbitTorque2020}.
Schemes allowing to quantify the number of magnons populating a given equilibrium or non-equilibrium mode thus are of key importance.

Ultrafast spectroscopy techniques have been instrumental in probing magnons, thereby providing time-resolved insights into magnetization dynamics ranging from femtosecond to picosecond timescales  \cite{kirilyukUltrafastOpticalManipulation2010}.
Most studies have focused on magnons in coherent response to external stimuli, such as microwave fields, ultrafast laser pulses, or THz radiation.
Incoherent dynamics also play a significant role in determining the overall magnetic behavior in thermal equilibrium and excited states (e.g. ultrafast demagnetization \cite{beaurepaireUltrafastSpinDynamics1996}), and are relevant in applications like heat-assisted magnetic recording \cite{hsuHeatassistedMagneticRecording2022,rottmayerHeatAssistedMagneticRecording2006}.
However, in typical ultrafast experiments those incoherent dynamics average out and are thus not directly accessible, although they influence the coherent response, e.g. governing ultrafast demagnetization \cite{beaurepaireUltrafastSpinDynamics1996,deCoherentIncoherentMagnons2024}, driving the generation of THz radiation \cite{zhangUltrafastTerahertzMagnetometry2020} or altering the energies of polaritons and excitons \cite{dirnbergerMagnetoopticsVanWaals2023}.

Investigations of incoherent magnon populations have primarily relied on frequency-domain techniques such as Brillouin light scattering (BLS) \cite{gubbiottiBrillouinLightScattering2010,wojewodaModelingMicrofocusedBrillouin2024,sandwegWiderangeWavevectorSelectivity2010,sebastianMicrofocusedBrillouinLight2015,demidovMagnonicWaveguidesStudied2015}, neutron scattering \cite{princepFullMagnonSpectrum2017}, or resonant inelastic X-ray scattering \cite{guObservingDifferentialSpin2025}, which lack ultrafast time resolution.
So far, a direct, quantitative, time-domain measurement of the magnon population is lacking.

In this study, we employ a time-domain technique -- femtosecond noise correlation spectroscopy (FemNoC) \cite{weissDiscoveryUltrafastSpontaneous2023,weissSubharmonicLockinDetection2024,weissQuantifyingAmplitudesUltrafast2025,weissFieldtuningUltrafastMagnetization2026} -- to concurrently probe both thermal magnon fluctuations, and magnons driven via ferromagnetic resonance in a magnetic garnet film.
Both populations can be distinguished and characterized by their respective time-domain waveforms and are quantitatively analyzed by modeling the magnon dispersion relation and the resulting magnetization correlation function.
This fact allows us to also quantitatively measure the number of magnons excited by a free-running microwave drive in the linear response regime, and beyond.
These results demonstrate that FemNoC allows filling the gap between frequency-domain techniques and ultrafast studies of coherent magnetization dynamics, providing direct access to the time-domain correlation function, which enables quantitative measurement of coherent and incoherent magnon populations in both thermal and excited states.

\begin{figure*}
    \includegraphics[width=\textwidth]{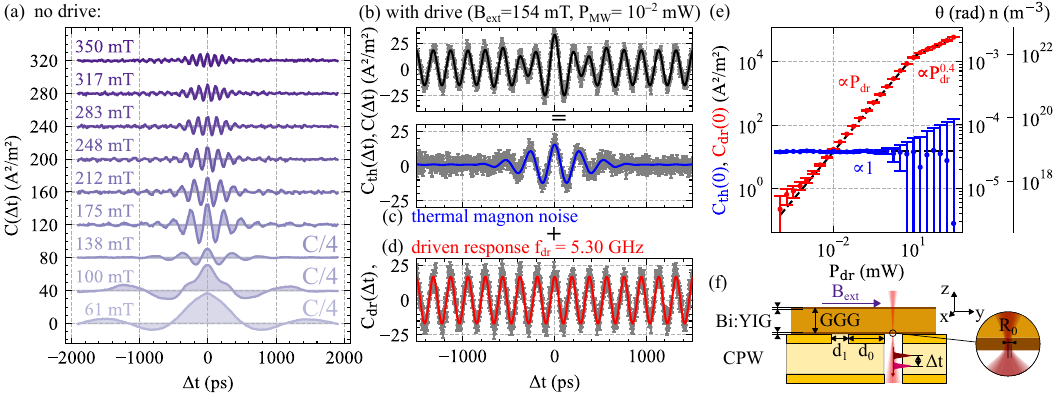}
    \caption{
        (a): Measured correlation amplitude $C(\Delta t)$ of the thermal magnon noise at selected $B_{\mathrm{ext}}$ and no microwave power applied ($P_{\mathrm{dr}}=0$).
        The curves are offset for clarity, with the bottom three scaled down by a factor of 4 for better visibility.
        (b-d): Measured $C(\Delta t)$ (gray) with a free-running microwave drive applied at $f_{\mathrm{dr}}=\qty{5.3}{\giga\hertz}$, and with $B_{\mathrm{ext}} = \qty{154}{\milli\tesla}$.
        The colored lines are fits to the respective waveforms.
        The full correlation function (b) can be decomposed into thermal magnons $C_{\mathrm{th}}$ (c) measured at $P_{\mathrm{dr}}=0$ and excited magnons $C_{\mathrm{dr}}$ (d).
        (e): Measurements at different $P_{\mathrm{dr}}$ with the amplitude of $C_{\mathrm{dr}}$ and $C_{\mathrm{th}}$ extracted by fitting.
        $C(0)$ can be converted to the precession cone angle $\theta_c$ and the corresponding magnon number per sample volume $n$ (right axis).
        The incoherent contribution remains constant, while the number of excited magnons $n$ increases linearly with the applied power $P_{\mathrm{dr}}$ until beginning to saturate at $n \approx 10^{22}\qty{}{\per\m\cubed}$.
        The respective scalings are indicated by black dashed lines.
        (f): Schematic of the setup.
        The sample is placed on a coplanar waveguide (CPW, $d_0 = \qty{1070}{\micro\m}$, $d_1=\qty{280}{\micro\m}$).
        The optical probes with delay time $\Delta t$ are transmitted in $z$ dir. through a hole in the dielectric medium.
        An external magnetic field $B_{\mathrm{ext}}$ is applied in the sample plane ($y$ dir.), perpendicular to the direction of the waveguide ($x$ dir.).
    }
    \label{fig:noise_correlation}
\end{figure*}

\begin{figure}
    \includegraphics[width=\columnwidth]{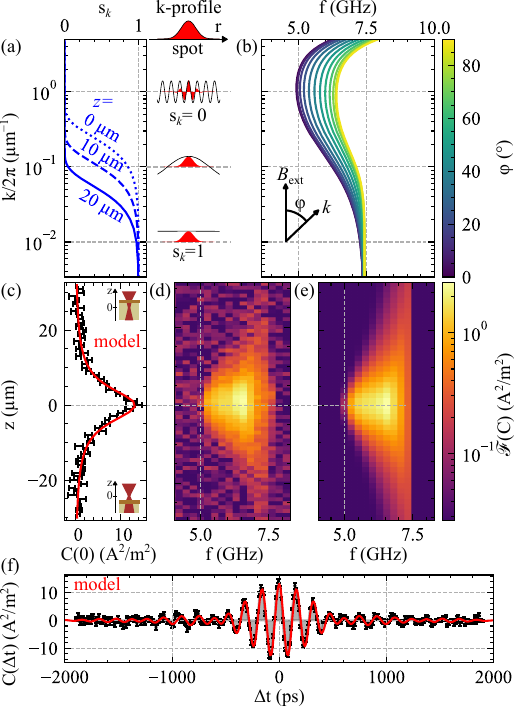}
    \caption{
        Analysis of the incoherent magnon signal at $B_{\mathrm{ext}}$ $= \qty{238}{\milli\tesla}$.
        (a) Simulated spot sensitivity to wave vectors $|\vec{k}|$ at different sample positions $z$.
        Comparison of the spatial profiles of the laser spot (red) and the magnon modes (black) for different $|\vec{k}|$.
        (b) Simulated magnon dispersion relation as a function of wave vector $\vec{k}$ and its direction with respect to the external magnetic field (colored lines).
        (c) Measured zero-delay correlation amplitude $C(0)$ as a function of the sample position $z$ along the optical axis (black dots) and simulated amplitude (red).
		(d) Frequency contributions of the measured correlation signal as a function of $z$ obtained by Fourier transforming the time-domain data.
        (e) Simulated frequency contributions to the correlation signal as a function of $z$, as determined from the convolution of the spot sensitivity and the dispersion relation.
        (f) Measured $C(\Delta t)$ at $z=\qty{0}{\micro\m}$ (black+gray) and the corresponding simulated signal (red).
    }
    \label{fig:zscananalysis}
\end{figure}

We investigate a $d^{\mathrm{z}} = \qty{2.5}{\micro\m}$ thick bismuth-substi\-tuted yttrium iron garnet (Bi:YIG, $\mathrm{Y_{1.6}Bi_{1.4}Fe_{5}O_{12}}$) film.
It is grown with the 111 direction out-of-plane via liquid phase epitaxy on a crystalline gadolinium gallium garnet substrate substituted with Ca, Mg, and Zr atoms (sGGG) to tailor the lattice constant.
Substituting yttrium with bismuth increases the magneto-optical response without significant disturbance of the magnetic properties \cite{yangStudyMagneticMagnetooptical2006}.
The sample is glued in flip-chip configuration onto a coplanar waveguide (CPW) as shown in \cref{fig:noise_correlation}\,f. 
An external magnetic field $B_{\mathrm{ext}}$ is applied in the sample plane ($y$ dir.), perpendicular to the waveguide ($x$ dir.).
The magnetization dynamics are then optically probed in transmission ($z$ dir.) by focusing the FemNoC probe pulses \cite{weissDiscoveryUltrafastSpontaneous2023} onto the sample through a hole in the dielectric medium of the waveguide.
The probe spot radius ($1/e^2$) is $R_0 = \qty{1.2(1)}{\micro\m}$ in focus and can be increased by moving the sample in $z$ dir. out of focus (see \cref{fig:zscananalysis}\,c).

FemNoC (see Refs.\,\cite{weissSubharmonicLockinDetection2024,weissQuantifyingAmplitudesUltrafast2025} for details) utilizes two linearly polarized laser pulses of central wavelengths of \qty{767}{\nano\m} and \qty{775}{\nano\m}, with an adjustable time delay $\Delta t$.
Upon transmission through the sample, the Faraday effect induces a polarization rotation proportional to the instantaneous magnetization component $M^{\mathrm{z}}$ along the beam direction.
We separate the two pulses by wavelength into individual balanced photodetection branches and subsequently perform electronic correlation.
Therefore, delay line length ($\pm\qty{1900}{\pico\s}$) and duration of the laser pulses (\qty{300}{\femto\s}) are the limiting factors for time resolution, allowing broadband detection of magnetization dynamics between \qty{0.52}{\giga\hertz} and \qty{1.6}{\tera\hertz}.
With appropriate calibration \cite{weissQuantifyingAmplitudesUltrafast2025}, FemNoC directly probes the magnetization correlation function \cite{weissSubharmonicLockinDetection2024}
\begin{align}
C(\Delta t) := \langle M^{\mathrm{z}}(t) M^{\mathrm{z}}(t + \Delta t) \rangle_t, \label{eq-1}
\end{align}
where $\langle\cdot\rangle_t$ denotes the average over time $t$.

The experiment is prepared by setting the external static field $B_{\mathrm{ext}}$ and applying the microwave drive to the CPW at frequency $f_{\mathrm{dr}}$ and power $P_{\mathrm{dr}}$.
All measurements are performed at room temperature.
We then record $C(\Delta t)$ by scanning the probe delay time $\Delta t$ averaging for \qty{1}{\second} per data point, revealing the respective magnetic fluctuations (\cref{fig:noise_correlation}).
To further increase the signal-to-noise ratio, we average over multiple individual scans.

We first discuss $C(\Delta t)$ in thermal equilibrium, without a microwave drive applied to the CPW.
\cref{fig:noise_correlation}\,a shows the measured correlation function at selected magnetic fields.
$C(\Delta t)$ peaks at zero delay and shows an oscillatory decay for increasing delay times.
As we will model later, this waveform is the sum of all thermally occupied magnon modes, each contributing a cosine-shaped correlation.
At low fields, $C(\Delta t)$ has a high amplitude and low oscillation frequency, while at high fields, the amplitude is significantly lower and the frequency higher.

We now compare this to the response under coherent microwave excitation.
Here, we drive the ferromagnetic resonance (FMR) with a free-running microwave applied to the CPW and measure $C(\Delta t)$ simultaneously.
\cref{fig:noise_correlation}\,b shows $C(\Delta t)$ recorded at $B_{\mathrm{ext}}=\qty{154}{\milli\tesla}$ for a fixed $f_{\mathrm{dr}}=\qty{5.3}{\giga\hertz}$. 
This pair of field and frequency fulfills the FMR condition.
For $P_{\mathrm{dr}}=\qty{-20}{\dBm}$, only a small increase in $C(\Delta t)$ in the same order of magnitude as the combined thermal magnon noise $C_\mathrm{th}$ (measured at $P_{\mathrm{dr}}=0$, \cref{fig:noise_correlation}\,c) is evident. 
To extract the corresponding coherent magnon contribution $C_{\mathrm{dr}}$, we subtract $C_{\mathrm{th}}$ from $C(\Delta t)$, obtaining the cosine shaped $C_{\mathrm{dr}}$ (\cref{fig:noise_correlation}\,d).

We now repeat this measurement as a function of $P_{\mathrm{dr}}$ and fit the data with the waveforms (colored lines in \cref{fig:noise_correlation}\,b-d.) determined above to extract the amplitudes $C_\mathrm{dr}(0)$ and $C_\mathrm{th}(0)$ as a function of $P_{\mathrm{dr}}$, shown in \cref{fig:noise_correlation}\,e.
In the investigated power range, we find the thermal magnon noise to remain unchanged within the degree of detectability, while the coherent amplitude increases linearly with the applied power up to a threshold of $P_{\mathrm{dr}} \approx \qty{10}{\milli\watt}$.
This is the expected behavior for linear response, assuming the microwave only overpopulates magnon modes with $k \approx 0$ (the Kittel mode), with the other modes unaffected \cite{stancilSpinWavesTheory2009}.
At higher driving powers, the amplitude of the coherent contribution scales sublinearly with roughly $P_{\mathrm{dr}}^{0.4}$
We attribute this saturation to the well-known onset of non-linear scattering \cite{suhlTheoryFerromagneticResonance1957}.
Since three-magnon scattering from the Kittel mode is prevented by absence of compatible modes, four-magnon scattering is the likely process at work.

We convert the measured correlation amplitude $C(0)$ to the precession cone angle $\theta_c = \arctan\left(\frac{1}{M_{\mathrm{sat}}}\sqrt{2 C(0)}\right)$ utilizing the calibration of the absolute amplitude of $C(\Delta t)$ \cite{weissQuantifyingAmplitudesUltrafast2025} and assuming circular precession of the magnetization.
The corresponding values are shown on the right of  \cref{fig:noise_correlation}\,e.
In order to quantitatively extract the magnon number per sample volume $n$ (\cref{eq:magnon_number}), we now model $C(\Delta t)$, taking into account the mode profile and thermal occupation of the magnon modes in the sample. 
The modeling approach utilizes similar ingredients as models describing Brillouin light scattering \cite{wojewodaModelingMicrofocusedBrillouin2024} but is fully quantitative.

In a first step, we express the detected magnetization signal $M^{\mathrm{z}}(t)$ as the convolution of the spatial magnon mode profiles $m^{\mathrm{z}}_{\vec{k}}(\vec r, z, t)$ \footnote{We denote variables as subscripts, e.g. $m^{\mathrm{z}}_{\vec{k}}$ as equivalent to $m^{\mathrm{z}}(\vec{k})$ for compactness in longer equations. To avoid confusion, we denote the labels for the $z$ component of the magnetization or thickness along $z$ as superscript $\mathrm{z}$.} with the Gaussian profile $I(\vec{r}, z)$ of the laser spots with radius $R_0$ and average power $P_{\mathrm{LS}}$.
Here, $\vec{r}$ is the in-plane coordinate, $z$ the out-of-plane coordinate of the sample, and $\vec{k}$ the in-plane wave vector of the magnon mode.
\begin{align}
    M^{\mathrm{z}} = \frac{1}{P_{\mathrm{LS}} d^{\mathrm{z}}}\int \mathrm{d}^2\vec{r}\int \mathrm{d}z \, I(|\vec{r}|,z) \sum_{\vec{k}} m^{\mathrm{z}}_{\vec{k}}(\vec{r}, z) \label{eq:beamconvolution}
\end{align}
We then express $m^{\mathrm{z}}_{\vec{k}}(\vec r, z, t) = a_{\vec{k}} \cos(\vec{k} \vec r - 2\pi f_{\vec{k}} t + \phi_{\vec{k}})$ as a plane wave \cite{stancilSpinWavesTheory2009} with the amplitude $a_{\vec{k}}$ and phase $\phi_{\vec{k}}$, with no $z$-profile as a first approximation.

The occupation of each mode in thermal equilibrium is given by the Bose-Einstein distribution $n_{\vec{k}} = (\exp(\frac{h f_{\vec{k}}}{k_B T}) - 1)^{-1}$ and describes the reduction of $M_{\mathrm{sat}}$ by $\frac{1}{V}\hbar\gamma n_{\vec{k}}$ in the direction of $B_{\mathrm{ext}}$.
To quantify $a_{\vec{k}}$ as a function $n_{\vec{k}}$, we assume each mode as low-amplitude circular precession with maintained total magnetization $M_\mathrm{sat}$.
Consequently, we get $M_{\mathrm{sat}}^2 = a_{\vec{k}}^2 + (M_{\mathrm{sat}} - \frac{1}{V}\hbar\gamma n_{\vec{k}})^2$, where $\gamma$ is the gyromagnetic ratio and $V$ the volume of the sample.
Neglecting terms of second order in $n_{\vec{k}}$, this simplifies to $a_{\vec{k}}^2 \approx 2 V^{-1} M_{\mathrm{sat}} \hbar\gamma n_{\vec{k}}$.

Calculating the convolution and correlation (see appendix for details) yields 
\begin{align}
    C(\Delta t) = \frac{M_\mathrm{sat} \hbar \gamma}{4 \pi^2 d^{\mathrm{z}}} \int \mathrm{d}^2 \vec{k} \, n_{\vec{k}} \cos(2\pi f_{\vec{k}} \Delta t) s_{\vec k}  \psi_{f} \psi_p \Psi. \label{eq:corrampl_final}
\end{align}
Here, $s_{\vec k} = \exp(-\vec{k}^2 R(z)^2/4)$ originates from \cref{eq:beamconvolution} and describes the sensitivity to different wave vectors $\vec{k}$ due to the laser spot size $R(z)$ at the sample position $z$, as sketched in \cref{fig:zscananalysis}\,a for different $z$-positions of the sample.
Modes with $\vec{k} \gg \frac{2}{R(z)}$ have multiple periods within the laser spot and thus average out.

As discussed in the appendix in detail, we account for a correction factor of $\psi_f$ for the frequency sensitivity of our measurement due to the effects of the laser repetition rate $f_{\mathrm{rep}}$.
Furthermore, we include the $z$ profile of the modes with the factor $\psi_p$.

In the following, we compare the measured correlation function of the thermal magnon background with the model described above.
We find a systematic deviation of $\Psi = \qty{0.86(25)}{}$ between the measured and simulated correlation amplitude, where the uncertainty is given by the compounded uncertainty of all prefactors in \cref{eq:corrampl_final}.
As $\Psi$ is in agreement with unity within the uncertainty, we conclude that our model faithfully predicts the thermal magnon correlation amplitude.
Given the fact, that our model ignores ellipticity we should expect $\Psi < 1$, as ellipticity reduces the $z$-component of the dynamic magnetization and thus the measured correlation amplitude (see detailed discussion in the appendix).

To simulate $C(\Delta t)$, for our specific sample, we calculate $f_{\vec{k}}$ at a given external magnetic field using the TetraX software package \cite{korberTetraXFiniteElementMicromagneticModeling2022,korberFiniteelementDynamicmatrixApproach2022a}.  
The result for $B_{\mathrm{ext}} = \qty{238}{\milli\tesla}$ is shown in \cref{fig:zscananalysis}\,b for modes with zero $z$ modulation as a function of the wave vector $|\vec{k}|$ and its direction with respect to the external field.
At small $|\vec{k}|$, the modes are dominated by dipolar interactions, while at larger $|\vec{k}|$, the exchange interaction dominates \cite{kalinikosTheoryDipoleexchangeSpin1986,harmsTheoryDipoleexchangeSpin2022}.
This leads to a minimum in the dispersion relation at $|\vec{k}| \approx \frac{1}{2\pi} \unit{\per\micro\m}$.

Scanning the sample through the laser focus in $z$ direction allows us to systematically change the radius $R(z)$ of the laser spot in the sample plane, which modifies the $\vec{k}$-space sensitivity $s_{\vec{k}}$ of our probing scheme as shown in \cref{fig:zscananalysis}\,a.
\cref{eq:corrampl_final} yields the expected frequency contributions to the correlation signal.  
Moving the sample out of focus decreases the sensitivity to higher wave vectors. 
This leads to a reduced overall correlation amplitude, as shown in \cref{fig:zscananalysis}\,c, where the simulation [\cref{eq:corrampl_final}] is plotted as a red line, in excellent agreement with the experimental data (black symbols).
Moreover, also the frequency contributions to the simulated correlation signal characteristically depend on spot size, as depicted in
\cref{fig:zscananalysis}\,e.

To compare this theoretical prediction with the experimental data, we Fourier-transform the measured correlation function for each $z$ position and obtain the frequency contributions shown in \cref{fig:zscananalysis}\,d.
The experimental data almost perfectly match the theoretical prediction in shape and amplitude.
As we discussed above, for the latter, we see a systematic deviation of $\Psi$ throughout all measurements, which we include in the theory curves for better comparability.
Also in the time domain, the simulated correlation function in focus fiducially reproduces the measured data (see \cref{fig:zscananalysis}\,f).
Finally, we extract the outline of the dispersion relation from the data measured in \cref{fig:zscananalysis}\,d, proving the backwards compatibility of our model (see details in the appendix).

\begin{figure}
    \includegraphics[width=\columnwidth]{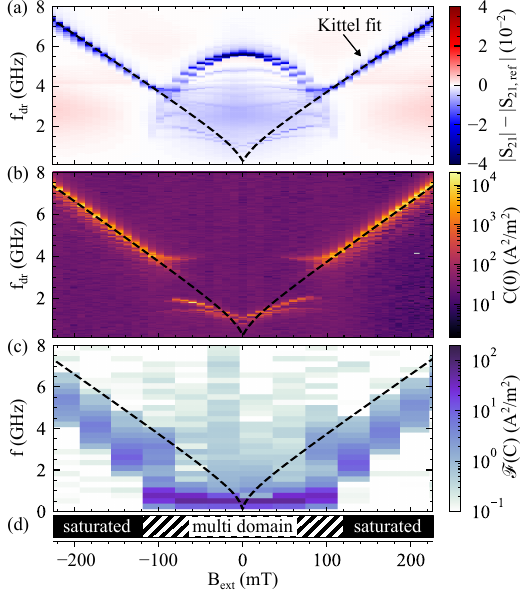}
    \caption{
        Comparison of the ferromagnetic resonance of the sample at $P_{\mathrm{dr}} = \qty{5}{\dBm}$ microwave field and the thermal magnon background.
        (a) Microwave transmission $|S_{21}|$ as a function of driving frequency $f_\mathrm{drive}$ and external field $B_{\mathrm{ext}}$ measured with a vector network analyzer minus the median off-resonant CPW transmission $|S_{21, \mathrm{ref}}|$.
        The resonance line in the single-domain state is fitted to the Kittel formula (black) and plotted in all three panels.
        (b) Zero-delay correlation amplitude $C(0)$ recorded simultaneously to (a).
        (c) Fourier transform $\mathcal{F}(C)$ of the unexcited magnon noise shown in \cref{fig:noise_correlation}\,b as a function of external field.
        The modes prominent in the thermal magnon background ($k>0$) are at lower frequencies than the modes excited by the microwave ($k\approx0$).
        (d) Magnetic state of the sample extracted from Faraday microscopy images (see SM).
    }
    \label{fig:fmr_comparison}
\end{figure}

Having established a quantitative model for the thermal magnon signal, we now turn to describing the signal under coherent excitation.
According to \cref{eq:corrampl_final}, $C(\Delta t)$ is also related to the total observable magnon number $n$ per sample volume
\begin{align}
    n = \frac{1}{V} \sum_{\vec{k}} n_{\vec{k}} s_{\vec{k}} = \frac{C(0)}{M_{\mathrm{sat}} \hbar \gamma \psi_p \Psi}. \label{eq:magnon_number}
\end{align}
Note that using the $\Psi$ factor determined from the thermal magnon background serves as an additional refinement of the systematic uncertainty of the absolute magnon number $n$, but is not necessary in case e.g. the thermal magnon background being below the detection threshold.
Thus, FemNoC provides a direct way to quantify the coherent magnon occupation in the sample under free-running microwave excitation without synchronization to the laser pulses, as other time-domain techniques require (e.g. \cite{dreyerSpinwaveLocalizationGuiding2021}).
The corresponding values for $n$ generated by coherent excitation are shown on the right axis of \cref{fig:noise_correlation}\,c.

In a final step, we discuss the response of the sample under microwave excitation as a function of frequency and field strength. 
To this end, we ramp the magnetic field between $\pm\qty{240}{\milli\tesla}$ in steps of $\approx\qty{10}{\milli\tesla}$, and for each field value sweep $\qty{0.1}{\giga\hertz} \le f_{\mathrm{dr}} \le \qty{7.9}{\giga\hertz}$ in steps of \qty{0.04}{\giga\hertz} at $P_{\mathrm{dr}}=\qty{5}{\dBm}$.
We simultaneously record the microwave transmission $|S_{21}(f_{\mathrm{dr}})|$ and $C(\Delta t=0)$.
The results are shown in \cref{fig:fmr_comparison}\,a and b together with the Fourier transform of $C(\Delta t)$ recorded without microwave drive in \cref{fig:fmr_comparison}\,c for comparison.

We find two regimes separated by a transition region around $\qty{120}{\milli\tesla}$.
At low fields, the absorption splits into multiple resonance lines, but only some of those lead to an increase of the observed correlation function (\cref{fig:fmr_comparison}\,b).
Due to the formation of magnetic domains in this region (see SM for details), a more complex resonance structure is expected, which we leave for future studies. 

For magnetic field magnitudes larger than \qty{120}{\milli\tesla}, a single microwave absorption peak in $|S_{21}|$ (\cref{fig:fmr_comparison}\,a)  coincides with a strong increase in the magnetization correlation amplitude $C(0)$ (\cref{fig:fmr_comparison}\,b).  
We attribute this resonance to the uniform precession mode of the magnetization at $\vec{k} \approx 0$ and fit the observed resonance frequencies with the Kittel formula for in-plane magnetized thin films \cite{kittelTheoryFerromagneticResonance1948}, shown as black dashed lines in \cref{fig:fmr_comparison} (see SM for details).
The excitation leads to a significantly increased population of the FMR mode compared to the thermal background, which becomes evident from the color scales of \cref{fig:fmr_comparison}\,b and c.
The absolute frequency values in the incoherent thermal spectrum are shifted to lower frequencies by about \qty{1.5}{\giga\hertz} compared to the data for finite $P_{\mathrm{dr}}$.
Additionally, the FMR line is significantly narrower than the width of the incoherent background.

Using the theoretical model presented above, these findings can be understood in terms of the dispersion relation of the magnon modes in the sample.
As the width of the CPW is significantly larger than the laser spot size, the microwave field excites magnon modes with $k\approx 0$, which are at about \qty{1.5}{\giga\hertz} higher frequencies than the minimum of the dispersion relation at finite $\vec{k}$-vectors the FemNoC experiment is sensitive to (see \cref{fig:zscananalysis}).
The signal in thermal equilibrium consists mainly of those finite-$\vec{k}$ magnons, which distribute over a broader frequency range dependent on the different directions of $\vec{k}$ with respect to the external field.
So, in addition to the incoherent magnon background, the FemNoC experiment is sensitive to the coherent $k \approx 0$ magnons excited by the microwave drive in a free-running measurement without external synchronization to the laser pulses. 
In fact, we have tacitly used this fact in the analysis of the coherent contribution in \cref{fig:noise_correlation}. 

To conclude, we have shown a direct and quantitative measurement of the magnon population in a magnetic garnet film using femtosecond noise correlation spectroscopy.
We present a quantitative magnon model that fiducially reproduces the incoherent magnon background measurements in the time domain and predicts the waveform and amplitude scaling of the incoherent magnon background based on a calculated dispersion relation and the laser spot size.
We also show that the $k$ sensitivity of FemNoC can be systematically tuned by changing the laser spot size.
The combination of our measurement technique with a microwave excitation then allows us to measure the coherent magnon response on top of the incoherent background and to extract the absolute magnon number density generated by the microwave excitation.
In the linear response regime, we find that the incoherent magnon background remains unaffected by the coherent drive, while the coherent magnon population scales linearly with the applied microwave power.
Furthermore, we quantify the onset of non-linear scattering at higher microwave powers when a magnon population of $n \approx 10^{22}\qty{}{\per\m\cubed}$ is reached.
Taken together, we have shown that FemNoC is able to supplement existing techniques for magnon spectroscopy, yielding direct access to the time domain and the absolute amplitudes of both incoherent and coherent magnon populations, which is especially relevant for ultrafast time scales and non-linear driving regimes.

The technique itself is not limited to magnonic systems, but can be generalized to a wide range of condensed matter systems.
As such our results pave the way for future experiments investigating incoherent dynamics in the time domain e.g., the spectrally resolved study of non-equilibrium dynamics on ultrafast timescales, or non-linear driving regimes.

\section*{Data availability}

The data shown in the figures of this study are made available on the scientific data repository KONDATA at \url{https://kondata.uni-konstanz.de/} with the identifier \texttt{PLACEHOLDER}.
Further information that support the findings of this study are available from the corresponding author upon reasonable request.

\section*{Acknowledgements}

We gratefully acknowledge discussions with the members of the \textit{SFB 1432}, in particular J.~Harms and W.~Belzig.
Computing resources were provided by the \textit{Scientific Compute Cluster in Konstanz (SCCKN)}.
This research was supported by the \textit{German Research Foundation (Deutsche Forschungsgemeinschaft, DFG)—Project No. 425217212 - SFB 1432}.

\bibliography{main}

@article{beaurepaireUltrafastSpinDynamics1996,
  title = {Ultrafast {{Spin Dynamics}} in {{Ferromagnetic Nickel}}},
  author = {Beaurepaire, E. and Merle, J.-C. and Daunois, A. and Bigot, J.-Y.},
  year = 1996,
  month = may,
  journal = {Physical Review Letters},
  volume = {76},
  number = {22},
  pages = {4250--4253},
  publisher = {American Physical Society},
  doi = {10.1103/PhysRevLett.76.4250},
  urldate = {2025-11-19}
}

@article{chumakAdvancesMagneticsRoadmap2022,
  title = {Advances in {{Magnetics Roadmap}} on {{Spin-Wave Computing}}},
  author = {Chumak, A. V. and Kabos, P. and Wu, M. and Abert, C. and Adelmann, C. and Adeyeye, A. O. and Akerman, J. and Aliev, F. G. and Anane, A. and Awad, A. and Back, C. H. and Barman, A. and Bauer, G. E. W. and Becherer, M. and Beginin, E. N. and Bittencourt, V. A. S. V. and Blanter, Y. M. and Bortolotti, P. and Boventer, I. and Bozhko, D. A. and Bunyaev, S. A. and Carmiggelt, J. J. and Cheenikundil, R. R. and Ciubotaru, F. and Cotofana, S. and Csaba, G. and Dobrovolskiy, O. V. and Dubs, C. and Elyasi, M. and Fripp, K. G. and Fulara, H. and Golovchanskiy, I. A. and {Gonzalez-Ballestero}, C. and Graczyk, P. and Grundler, D. and Gruszecki, P. and Gubbiotti, G. and Guslienko, K. and Haldar, A. and Hamdioui, S. and Hertel, R. and Hillebrands, B. and Hioki, T. and Houshang, A. and Hu, C.-M. and Huebl, H. and Huth, M. and Iacocca, E. and Jungfleisch, M. B. and Kakazei, G. N. and Khitun, A. and Khymyn, R. and Kikkawa, T. and Klaui, M. and Klein, O. and Klos, J. W. and Knauer, S. and Koraltan, S. and Kostylev, M. and Krawczyk, M. and Krivorotov, I. N. and Kruglyak, V. V. and {Lachance-Quirion}, D. and Ladak, S. and Lebrun, R. and Li, Y. and Lindner, M. and Macedo, R. and Mayr, S. and Melkov, G. A. and Mieszczak, S. and Nakamura, Y. and Nembach, H. T. and Nikitin, A. A. and Nikitov, S. A. and Novosad, V. and Otalora, J. A. and Otani, Y. and Papp, A. and Pigeau, B. and Pirro, P. and Porod, W. and Porrati, F. and Qin, H. and Rana, B. and Reimann, T. and Riente, F. and {Romero-Isart}, O. and Ross, A. and Sadovnikov, A. V. and Safin, A. R. and Saitoh, E. and Schmidt, G. and Schultheiss, H. and Schultheiss, K. and Serga, A. A. and Sharma, S. and Shaw, J. M. and Suess, D. and Surzhenko, O. and Szulc, K. and Taniguchi, T. and Urbanek, M. and Usami, K. and Ustinov, A. B. and Van Der Sar, T. and Van Dijken, S. and Vasyuchka, V. I. and Verba, R. and Kusminskiy, S. Viola and Wang, Q. and Weides, M. and Weiler, M. and Wintz, S. and Wolski, S. P. and Zhang, X.},
  year = 2022,
  month = jun,
  journal = {IEEE Transactions on Magnetics},
  volume = {58},
  number = {6},
  pages = {1--72},
  issn = {0018-9464, 1941-0069},
  doi = {10.1109/TMAG.2022.3149664},
  urldate = {2025-11-27},
  copyright = {https://creativecommons.org/licenses/by/4.0/legalcode}
}

@article{chumakMagnonSpintronics2015,
  title = {Magnon Spintronics},
  author = {Chumak, A. V. and Vasyuchka, V. I. and Serga, A. A. and Hillebrands, B.},
  year = 2015,
  month = jun,
  journal = {Nature Physics},
  volume = {11},
  number = {6},
  pages = {453--461},
  publisher = {Nature Publishing Group},
  issn = {1745-2481},
  doi = {10.1038/nphys3347},
  urldate = {2024-06-27},
  copyright = {2014 Springer Nature Limited},
  langid = {english}
}

@article{deCoherentIncoherentMagnons2024,
  title = {Coherent and Incoherent Magnons Induced by Strong Ultrafast Demagnetization in Thin Permalloy Films},
  author = {De, Anulekha and Lentfert, Akira and Scheuer, Laura and Stadtm{\"u}ller, Benjamin and {von Freymann}, Georg and Aeschlimann, Martin and Pirro, Philipp},
  year = 2024,
  month = jan,
  journal = {Physical Review B},
  volume = {109},
  number = {2},
  pages = {024422},
  publisher = {American Physical Society},
  doi = {10.1103/PhysRevB.109.024422},
  urldate = {2026-08-06}
}

@article{demidovMagnonicWaveguidesStudied2015,
  title = {Magnonic {{Waveguides Studied}} by {{Microfocus Brillouin Light Scattering}}},
  author = {Demidov, Vladislav E. and Demokritov, Sergej O.},
  year = 2015,
  month = apr,
  journal = {IEEE Transactions on Magnetics},
  volume = {51},
  number = {4},
  pages = {1--15},
  issn = {0018-9464, 1941-0069},
  doi = {10.1109/TMAG.2014.2388196},
  urldate = {2025-01-30},
  copyright = {https://ieeexplore.ieee.org/Xplorehelp/downloads/license-information/IEEE.html},
  langid = {english}
}

@article{demidovSpinOrbitTorque2020,
  title = {Spin--Orbit-Torque Magnonics},
  author = {Demidov, V. E. and Urazhdin, S. and Anane, A. and Cros, V. and Demokritov, S. O.},
  year = 2020,
  month = may,
  journal = {Journal of Applied Physics},
  volume = {127},
  number = {17},
  pages = {170901},
  issn = {0021-8979},
  doi = {10.1063/5.0007095},
  urldate = {2025-11-27}
}

@article{dirnbergerMagnetoopticsVanWaals2023,
  title = {Magneto-Optics in a van Der {{Waals}} Magnet Tuned by Self-Hybridized Polaritons},
  author = {Dirnberger, Florian and Quan, Jiamin and Bushati, Rezlind and Diederich, Geoffrey M. and Florian, Matthias and Klein, Julian and Mosina, Kseniia and Sofer, Zdenek and Xu, Xiaodong and Kamra, Akashdeep and {Garc{\'i}a-Vidal}, Francisco J. and Al{\`u}, Andrea and Menon, Vinod M.},
  year = 2023,
  month = aug,
  journal = {Nature},
  volume = {620},
  number = {7974},
  pages = {533--537},
  publisher = {Nature Publishing Group},
  issn = {1476-4687},
  doi = {10.1038/s41586-023-06275-2},
  urldate = {2026-08-06},
  copyright = {2023 The Author(s), under exclusive licence to Springer Nature Limited},
  langid = {english}
}

@article{dreyerSpinwaveLocalizationGuiding2021,
  title = {Spin-Wave Localization and Guiding by Magnon Band Structure Engineering in Yttrium Iron Garnet},
  author = {Dreyer, Rouven and Liebing, Niklas and Edwards, Eric R. J. and M{\"u}ller, Andreas and Woltersdorf, Georg},
  year = 2021,
  month = jun,
  journal = {Physical Review Materials},
  volume = {5},
  number = {6},
  pages = {064411},
  publisher = {American Physical Society},
  doi = {10.1103/PhysRevMaterials.5.064411},
  urldate = {2026-02-24}
}

@article{flebus2024MagnonicsRoadmap2024,
  title = {The 2024 Magnonics Roadmap},
  author = {Flebus, Benedetta and Grundler, Dirk and Rana, Bivas and Otani, YoshiChika and Barsukov, Igor and Barman, Anjan and Gubbiotti, Gianluca and Landeros, Pedro and Akerman, Johan and Ebels, Ursula and Pirro, Philipp and Demidov, Vladislav E and Schultheiss, Katrin and Csaba, Gyorgy and Wang, Qi and Ciubotaru, Florin and Nikonov, Dmitri E and Che, Ping and Hertel, Riccardo and Ono, Teruo and Afanasiev, Dmytro and Mentink, Johan and Rasing, Theo and Hillebrands, Burkard and Kusminskiy, Silvia Viola and Zhang, Wei and Du, Chunhui Rita and Finco, Aurore and {van der Sar}, Toeno and Luo, Yunqiu Kelly and Shiota, Yoichi and Sklenar, Joseph and Yu, Tao and Rao, Jinwei},
  year = 2024,
  month = jun,
  journal = {Journal of Physics: Condensed Matter},
  volume = {36},
  number = {36},
  pages = {363501},
  publisher = {IOP Publishing},
  issn = {0953-8984},
  doi = {10.1088/1361-648X/ad399c},
  urldate = {2025-11-27},
  langid = {english}
}

@article{gubbiottiBrillouinLightScattering2010,
  title = {Brillouin Light Scattering Studies of Planar Metallic Magnonic Crystals},
  author = {Gubbiotti, G and Tacchi, S and Madami, M and Carlotti, G and Adeyeye, A O and Kostylev, M},
  year = 2010,
  month = jun,
  journal = {Journal of Physics D: Applied Physics},
  volume = {43},
  number = {26},
  pages = {264003},
  issn = {0022-3727},
  doi = {10.1088/0022-3727/43/26/264003},
  urldate = {2025-11-27},
  langid = {english}
}

@article{guObservingDifferentialSpin2025,
  title = {Observing Differential Spin Currents by Resonant Inelastic {{X-ray}} Scattering},
  author = {Gu, Yanhong and Barker, Joseph and Li, Jiemin and Kikkawa, Takashi and Camino, Fernando and Kisslinger, Kim and Sinsheimer, John and Lienhard, Lukas and Bauer, Jackson J. and Ross, Caroline A. and Basov, Dmitri N. and Saitoh, Eiji and Pelliciari, Jonathan and Bauer, Gerrit E. W. and Bisogni, Valentina},
  year = 2025,
  month = sep,
  journal = {Nature},
  volume = {645},
  number = {8082},
  pages = {900--905},
  publisher = {Nature Publishing Group},
  issn = {1476-4687},
  doi = {10.1038/s41586-025-09488-9},
  urldate = {2025-11-27},
  copyright = {2025 This is a U.S. Government work and not under copyright protection in the US; foreign copyright protection may apply},
  langid = {english}
}

@article{harmsTheoryDipoleexchangeSpin2022,
  title = {Theory of the Dipole-Exchange Spin Wave Spectrum in Ferromagnetic Films with in-Plane Magnetization Revisited},
  author = {Harms, J. S. and Duine, R. A.},
  year = 2022,
  month = sep,
  journal = {Journal of Magnetism and Magnetic Materials},
  volume = {557},
  pages = {169426},
  issn = {0304-8853},
  doi = {10.1016/j.jmmm.2022.169426},
  urldate = {2025-11-27}
}

@book{hechtOptik2018,
  title = {{Optik}},
  author = {Hecht, Eugene},
  year = 2018,
  month = mar,
  publisher = {De Gruyter},
  doi = {10.1515/9783110526653},
  urldate = {2023-10-06},
  copyright = {De Gruyter expressly reserves the right to use all content for commercial text and data mining within the meaning of Section 44b of the German Copyright Act.},
  isbn = {978-3-11-052665-3},
  langid = {ngerman}
}

@article{hsuHeatassistedMagneticRecording2022,
  title = {Heat-Assisted Magnetic Recording --- {{Micromagnetic}} Modeling of Recording Media and Areal Density: {{A}} Review},
  shorttitle = {Heat-Assisted Magnetic Recording --- {{Micromagnetic}} Modeling of Recording Media and Areal Density},
  author = {Hsu, Wei-Heng and Victora, R. H.},
  year = 2022,
  month = dec,
  journal = {Journal of Magnetism and Magnetic Materials},
  volume = {563},
  pages = {169973},
  issn = {0304-8853},
  doi = {10.1016/j.jmmm.2022.169973},
  urldate = {2025-11-27}
}

@article{kalinikosTheoryDipoleexchangeSpin1986,
  title = {Theory of Dipole-Exchange Spin Wave Spectrum for Ferromagnetic Films with Mixed Exchange Boundary Conditions},
  author = {Kalinikos, B. A. and Slavin, A. N.},
  year = 1986,
  month = dec,
  journal = {Journal of Physics C: Solid State Physics},
  volume = {19},
  number = {35},
  pages = {7013},
  issn = {0022-3719},
  doi = {10.1088/0022-3719/19/35/014},
  urldate = {2025-01-30},
  langid = {english}
}

@article{kirilyukUltrafastOpticalManipulation2010,
  title = {Ultrafast Optical Manipulation of Magnetic Order},
  author = {Kirilyuk, Andrei and Kimel, Alexey V. and Rasing, Theo},
  year = 2010,
  month = sep,
  journal = {Reviews of Modern Physics},
  volume = {82},
  number = {3},
  pages = {2731--2784},
  publisher = {American Physical Society},
  doi = {10.1103/RevModPhys.82.2731},
  urldate = {2025-11-24}
}

@article{kittelTheoryFerromagneticResonance1948,
  title = {On the {{Theory}} of {{Ferromagnetic Resonance Absorption}}},
  author = {Kittel, Charles},
  year = 1948,
  month = jan,
  journal = {Physical Review},
  volume = {73},
  number = {2},
  pages = {155--161},
  publisher = {American Physical Society},
  doi = {10.1103/PhysRev.73.155},
  urldate = {2024-06-26}
}

@article{korberFiniteelementDynamicmatrixApproach2022a,
  title = {Finite-Element Dynamic-Matrix Approach for Propagating Spin Waves: {{Extension}} to Mono- and Multi-Layers of Arbitrary Spacing and Thickness},
  shorttitle = {Finite-Element Dynamic-Matrix Approach for Propagating Spin Waves},
  author = {K{\"o}rber, L. and Hempel, A. and Otto, A. and Gallardo, R. A. and Henry, Y. and Lindner, J. and K{\'a}kay, A.},
  year = 2022,
  month = nov,
  journal = {AIP Advances},
  volume = {12},
  number = {11},
  pages = {115206},
  issn = {2158-3226},
  doi = {10.1063/5.0107457},
  urldate = {2025-05-08}
}

@misc{korberTetraXFiniteElementMicromagneticModeling2022,
  title = {{{TetraX}}: {{Finite-Element Micromagnetic-Modeling Package}}},
  shorttitle = {{{TetraX}}},
  author = {K{\"o}rber, Lukas and Quasebarth, Gwendolyn and Hempel, Alexander and Zahn, Friedrich and Otto, Andreas and Westphal, Elmar and Hertel, Riccardo and Kakay, Attila},
  year = 2022,
  month = jan,
  doi = {10.14278/RODARE.1418},
  urldate = {2025-03-07},
  copyright = {GNU General Public License v3.0 only, Open Access},
  howpublished = {Rodare}
}

@article{krizakovaSpinorbitTorqueSwitching2022,
  title = {Spin-Orbit Torque Switching of Magnetic Tunnel Junctions for Memory Applications},
  author = {Krizakova, Viola and Perumkunnil, Manu and Couet, S{\'e}bastien and Gambardella, Pietro and Garello, Kevin},
  year = 2022,
  month = nov,
  journal = {Journal of Magnetism and Magnetic Materials},
  volume = {562},
  pages = {169692},
  issn = {0304-8853},
  doi = {10.1016/j.jmmm.2022.169692},
  urldate = {2025-11-27}
}

@article{manchonCurrentinducedSpinorbitTorques2019,
  title = {Current-Induced Spin-Orbit Torques in Ferromagnetic and Antiferromagnetic Systems},
  author = {Manchon, A. and {\v Z}elezn{\'y}, J. and Miron, I. M. and Jungwirth, T. and Sinova, J. and Thiaville, A. and Garello, K. and Gambardella, P.},
  year = 2019,
  month = sep,
  journal = {Reviews of Modern Physics},
  volume = {91},
  number = {3},
  pages = {035004},
  publisher = {American Physical Society},
  doi = {10.1103/RevModPhys.91.035004},
  urldate = {2025-11-27}
}

@article{princepFullMagnonSpectrum2017,
  title = {The Full Magnon Spectrum of Yttrium Iron Garnet},
  author = {Princep, Andrew J. and Ewings, Russell A. and Ward, Simon and T{\'o}th, Sandor and Dubs, Carsten and Prabhakaran, Dharmalingam and Boothroyd, Andrew T.},
  year = 2017,
  month = nov,
  journal = {npj Quantum Materials},
  volume = {2},
  number = {1},
  pages = {1--5},
  publisher = {Nature Publishing Group},
  issn = {2397-4648},
  doi = {10.1038/s41535-017-0067-y},
  urldate = {2024-05-09},
  copyright = {2017 The Author(s)},
  langid = {english}
}

@article{rottmayerHeatAssistedMagneticRecording2006,
  title = {Heat-{{Assisted Magnetic Recording}}},
  author = {Rottmayer, R.E. and Batra, S. and Buechel, D. and Challener, W.A. and Hohlfeld, J. and Kubota, Y. and Li, L. and Lu, B. and Mihalcea, C. and Mountfield, K. and Pelhos, K. and Peng, C. and Rausch, T. and Seigler, M.A. and Weller, D. and Yang, X.-M.},
  year = 2006,
  month = oct,
  journal = {IEEE Transactions on Magnetics},
  volume = {42},
  number = {10},
  pages = {2417--2421},
  issn = {1941-0069},
  doi = {10.1109/TMAG.2006.879572},
  urldate = {2025-11-27}
}

@article{sandwegWiderangeWavevectorSelectivity2010,
  title = {Wide-Range Wavevector Selectivity of Magnon Gases in {{Brillouin}} Light Scattering Spectroscopy},
  author = {Sandweg, C. W. and Jungfleisch, M. B. and Vasyuchka, V. I. and Serga, A. A. and Clausen, P. and Schultheiss, H. and Hillebrands, B. and Kreisel, A. and Kopietz, P.},
  year = 2010,
  month = jul,
  journal = {Review of Scientific Instruments},
  volume = {81},
  number = {7},
  pages = {073902},
  issn = {0034-6748},
  doi = {10.1063/1.3454918},
  urldate = {2025-11-27}
}

@article{sebastianMicrofocusedBrillouinLight2015,
  title = {Micro-Focused {{Brillouin}} Light Scattering: Imaging Spin Waves at the Nanoscale},
  shorttitle = {Micro-Focused {{Brillouin}} Light Scattering},
  author = {Sebastian, Thomas and Schultheiss, Katrin and Obry, Bj{\~A}{\P}rn and Hillebrands, Burkard and Schultheiss, Helmut},
  year = 2015,
  month = jun,
  journal = {Frontiers in Physics},
  volume = {3},
  issn = {2296-424X},
  doi = {10.3389/fphy.2015.00035},
  urldate = {2025-01-30},
  langid = {english}
}

@book{stancilSpinWavesTheory2009,
  title = {Spin {{Waves}}: {{Theory}} and {{Applications}}},
  shorttitle = {Spin {{Waves}}},
  author = {Stancil, Daniel D. and Prabhakar, Anil},
  year = 2009,
  publisher = {Springer US},
  address = {Boston, MA},
  doi = {10.1007/978-0-387-77865-5},
  urldate = {2023-08-27},
  isbn = {978-0-387-77864-8 978-0-387-77865-5},
  langid = {english}
}

@article{suhlTheoryFerromagneticResonance1957,
  title = {The Theory of Ferromagnetic Resonance at High Signal Powers},
  author = {Suhl, H.},
  year = 1957,
  month = jan,
  journal = {Journal of Physics and Chemistry of Solids},
  volume = {1},
  number = {4},
  pages = {209--227},
  issn = {0022-3697},
  doi = {10.1016/0022-3697(57)90010-0},
  urldate = {2026-08-05}
}

@article{weissDiscoveryUltrafastSpontaneous2023,
  title = {Discovery of Ultrafast Spontaneous Spin Switching in an Antiferromagnet by Femtosecond Noise Correlation Spectroscopy},
  author = {Weiss, M. A. and Herbst, A. and Schlegel, J. and Dannegger, T. and Evers, M. and Donges, A. and Nakajima, M. and Leitenstorfer, A. and Goennenwein, S. T. B. and Nowak, U. and Kurihara, T.},
  year = 2023,
  month = nov,
  journal = {Nature Communications},
  volume = {14},
  number = {1},
  pages = {7651},
  publisher = {Nature Publishing Group},
  issn = {2041-1723},
  doi = {10.1038/s41467-023-43318-8},
  urldate = {2023-11-29},
  copyright = {2023 The Author(s)},
  langid = {english}
}

@article{weissFieldtuningUltrafastMagnetization2026,
  title = {Field-Tuning of Ultrafast Magnetization Fluctuations in $\mathrm{Sm_{0.7}Er_{0.3}FeO_{3}}$},
  author = {Weiss, Marvin Alexander and Schlegel, Julius and Anic, Daniel and Steiner, Emil and Herbst, Franz Stefan and Nakajima, Makoto and Kurihara, Takayuki and Leitenstorfer, Alfred and Nowak, Ulrich and Goennenwein, Sebastian T. B.},
  year = 2026,
  month = mar,
  journal = {Physical Review B},
  issn = {2469-9950, 2469-9969},
  doi = {10.1103/sd5w-n7fw},
  urldate = {2026-04-01},
  langid = {english}
}

@article{weissQuantifyingAmplitudesUltrafast2025,
  title = {Quantifying the Amplitudes of Ultrafast Magnetization Fluctuations in $\mathrm{Sm_{0.7}Er_{0.3}FeO_{3}}$ Using Femtosecond Noise-Correlation Spectroscopy},
  author = {Weiss, M.A. and Herbst, F.S. and Skobjin, G. and Eggert, S. and Nakajima, M. and Reustlen, D. and Leitenstorfer, A. and Goennenwein, S.T.B. and Kurihara, T.},
  year = 2025,
  month = oct,
  journal = {Physical Review Applied},
  volume = {24},
  number = {4},
  pages = {044021},
  publisher = {American Physical Society},
  doi = {10.1103/wkmb-ddwv},
  urldate = {2025-10-16}
}

@article{weissSubharmonicLockinDetection2024,
  title = {Subharmonic Lock-in Detection and Its Optimization for Femtosecond Noise Correlation Spectroscopy},
  author = {Weiss, M. A. and Herbst, F. S. and Eggert, S. and Nakajima, M. and Leitenstorfer, A. and Goennenwein, S. T. B. and Kurihara, T.},
  year = 2024,
  month = aug,
  journal = {Review of Scientific Instruments},
  volume = {95},
  number = {8},
  pages = {083005},
  issn = {0034-6748},
  doi = {10.1063/5.0208499},
  urldate = {2024-08-21}
}

@article{wojewodaModelingMicrofocusedBrillouin2024,
  title = {Modeling of Microfocused {{Brillouin}} Light Scattering Spectra},
  author = {Wojewoda, Ond{\v r}ej and Hrto{\v n}, Martin and Urb{\'a}nek, Michal},
  year = 2024,
  month = dec,
  journal = {Physical Review B},
  volume = {110},
  number = {22},
  pages = {224428},
  publisher = {American Physical Society},
  doi = {10.1103/PhysRevB.110.224428},
  urldate = {2025-11-27}
}

@article{yangStudyMagneticMagnetooptical2006,
  title = {Study of Magnetic and Magneto-Optical Properties of Heavily Doped Bismuth Substitute Yttrium Iron Garnet ({{Bi}}:{{YIG}}) Film},
  shorttitle = {Study of Magnetic and Magneto-Optical Properties of Heavily Doped Bismuth Substitute Yttrium Iron Garnet ({{Bi}}},
  author = {Yang, Qinghui and Zhang, Huaiwu and Liu, Yingli},
  year = 2006,
  month = oct,
  journal = {Rare Metals},
  volume = {25},
  number = {6, Supplement 1},
  pages = {557--561},
  issn = {1001-0521},
  doi = {10.1016/S1001-0521(07)60145-4},
  urldate = {2024-05-10}
}

@article{zhangUltrafastTerahertzMagnetometry2020,
  title = {Ultrafast Terahertz Magnetometry},
  author = {Zhang, Wentao and Maldonado, Pablo and Jin, Zuanming and Seifert, Tom S. and Arabski, Jacek and Schmerber, Guy and Beaurepaire, Eric and Bonn, Mischa and Kampfrath, Tobias and Oppeneer, Peter M. and Turchinovich, Dmitry},
  year = 2020,
  month = aug,
  journal = {Nature Communications},
  volume = {11},
  number = {1},
  pages = {4247},
  publisher = {Nature Publishing Group},
  issn = {2041-1723},
  doi = {10.1038/s41467-020-17935-6},
  urldate = {2026-08-06},
  copyright = {2020 The Author(s)},
  langid = {english}
}

\section*{Appendix}
\appendix

\subsection{Determining the dominant k-value per frequency} \label{subsec:dominantk}

In \cref{fig:zscananalysis}\,d, we show the frequency contributions to the correlation signal as a function of the sample position $z$.
We have compared this data to the theoretical prediction in \cref{fig:zscananalysis}\,e, which is calculated from the convolution of the spot sensitivity and the dispersion relation.
A more detailed analysis of the data in \cref{fig:zscananalysis}\,d also allows us to revert this analysis and directly extract the dominant wave vector contribution $|\vec{k}|$ for each frequency bin.

\begin{figure}[b]
    \centering
    \includegraphics[width=\columnwidth]{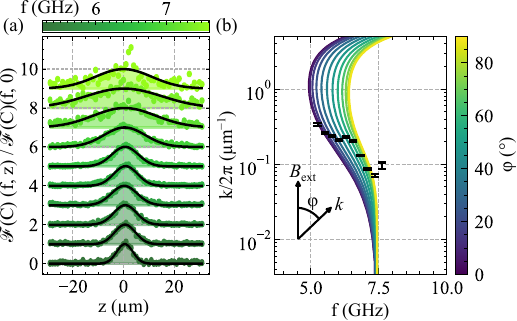}
    \caption{
        (a) Normalized amplitude of the frequency contributions of the correlation amplitude as a function of sample position $z$ at $B_{\mathrm{ext}} = \qty{239}{\milli \tesla}$.
        Each curve is fitted with the $k$ sensitivity $s(\vec{k})$ from \cref{eq:corrampl_final} to extract the dominant $|\vec{k}|$ value for each frequency bin.
        (b) Extracted dominant $|\vec{k}|$ values for each frequency bin (black dots) and the calculated dispersion relation (colored lines) from \cref{fig:zscananalysis}\,b.
    }
    \label{fig:zscan_kfit}
\end{figure}

To achieve this, we normalize the amplitudes of each frequency bin in \cref{fig:zscananalysis}\,d and fit it to the $k$ sensitivity $s(\vec{k})$ from \cref{eq:corrampl_final}, as shown in \cref{fig:zscan_kfit}\,a.
The extracted k-value represents the most dominant magnon mode of each frequency bin.
Within the sensitivity range of the setup, this is the mode with the highest $|\vec{k}|$ value at this frequency.
We display these values in \cref{fig:zscan_kfit}\,b and find good agreement with the calculated dispersion relation of the sample (same as in \cref{fig:zscananalysis}\,b), which proves the backwards compatibility of our model.

\subsection{Detailed derivation of the magnon model}

In the main text, we have presented the idea of our fitting model and the final result for the correlation function $C(\Delta t)$.
In the following, we outline the detailed derivation of the model, which is based on the convolution of the magnon mode profiles with the laser spot profile and subsequent calculation of the correlation function.

\paragraph{Beam profile:} We model the beam as Gaussian with radius $R$ (intensity $1/e^{2}$) as a function of position along the $z$ axis
\begin{align}
R(z) = R_{0} \sqrt{1+\left( \frac{\lambda (z-z_{f})}{\pi R_{0}^2 n} \right)^2}.
\end{align}
This results in the intensity distribution $I$ as a function of the distance $r$ from the beam axis
\begin{align}
I(r,z) = \frac{2P_{\mathrm{LS}}}{\pi R(z)^2} \exp\left(-2\frac{r^2}{R(z)^2}\right)
\end{align}
Here $R_{0}$ is the radius in focus, $P_{\mathrm{LS}}$ the total power of the beam, $\lambda$ the wavelength of the light, and $z_f$ the position of the focus relative to the sample position ($z=0$)\cite{hechtOptik2018}.

\paragraph{Magnon mode:} As a first-order approximation, we assume a magnon mode of the form of a plane wave with flat $z$ profile
\begin{align}
    m^{\mathrm{z}}_{\vec{k}}(\vec{r}, z) = a_{\vec{k}} \cos(\vec{k}\vec{r} - 2\pi f_{\vec{k}}t + \phi_{\vec{k}}) \cdot\mathrm{rect}\left( \frac{z}{d^{z}}  \right).
\end{align}
Note that in our coordinate system the magnetic field points in the $y$ direction.
Here, $\vec{r}=(x,y)$ and $\vec{k}=(k_x,k_y)$ are the two-dimensional, in-plane position and wave vectors, $\mathrm{rect}$ is the rectangular function, $d^{\mathrm{z}}$ the thickness of the sample, $a_{\vec{k}}$ the amplitude of the mode, $f_{\vec{k}}$ its frequency, as given by the dispersion relation, and $\phi_{\vec{k}}$ the phase of the mode.
To include effects of non-flat modes, we can replace the rectangular function with their respective $z$ profiles.

We quantify the precession amplitude $a_{\vec{k}}$ by assuming circular precession for each mode, resulting in
\begin{align}
    M_{s}^2 = a_{\vec{k}}^2 + \left( M_{s} - \frac{\hbar\gamma}{V} n_{\vec{k}} \right)^2 \Rightarrow a_{\vec{k}} \approx \frac{2 M_{s} \hbar \gamma}{V} n_{\vec{k}} \label{eq:circ_ampl_estimation}
\end{align}
where $M_{s}$ is the saturation magnetization, $\gamma$ the gyromagnetic ratio, $V$ the sample volume, and $n_{\vec{k}}$ the magnon occupation number.

\paragraph{Observed Magnetization:} We obtain the magnetization observed by the laser pulses $M^{\mathrm{z}}$ by overlapping their intensity distribution and the magnon mode profile as given in \cref{eq:beamconvolution}.
We solve this to
\begin{align}
    M^{\mathrm{z}} = \sum_{\vec{k}} a_{\vec{k}} \cos(2\pi f_{\vec{k}} t - \phi_{\vec{k}}) \exp\left( -\frac{1}{8}k^2 R(z)^2 \right)
\end{align}

\paragraph{Correlation:}
We now calculate the correlation function, which averages out the phases $\phi_{\vec{k}}$ and yields
\begin{align}
    C(\Delta t) &=\langle M^{\mathrm{z}}(t) M^{\mathrm{z}} (t+\Delta t) \rangle_{t} \\
    &= \sum_{\vec{k}} \frac{a_{\vec{k}}^2}{2} \cos(2\pi f_{\vec{k}} \Delta t) \exp\left( -\frac{k^2 R_{0}^2}{4} \right). \label{eq:corramp_sum}
\end{align}
This simplifies to \cref{eq:corrampl_final} when approximating the dispersion relation as continuous.

\subsection{Additional Effects:}
\begin{table*}
    \centering
    \begin{tabular}{l|l|r|l}
         & Value & rel. uncertainty & Origin \\ \hline
         $C_{\alpha, \mathrm{ave}}$ & \qty{4.60(77)}{\micro\V\per(\micro\rad)\squared} & \qty{17}{\%} & setup specific factor (see \cite{weissQuantifyingAmplitudesUltrafast2025}) \\
         $C_{M}$ & \qty{0.56(8)}{(\micro\rad)\squared\per(\A\per\m)\squared} & \qty{14}{\%} & sample specific factor derived like in \cite{weissQuantifyingAmplitudesUltrafast2025} \\
         $M_{\mathrm{sat}}$& \qty{145(5)}{\kilo\A\per\m} & \qty{3.5}{\%} & magnetometry measurements \\
         $\gamma$ & \qty{1.76(3)e11}{\per\T\per\s} & \qty{1.7}{\%} & FMR fit \\
         $d^{\mathrm{z}}$ & \qty{2.5(3)}{\micro\m} & \qty{12}{\%} & Faraday rotation + SEM imaging \\
         $\psi_{f}$ & \qty{0.127(5)}{} & \qty{3.9}{\%} & coherent response measurement (see SM)\\
         $\psi_{\mathrm{p}}$ & \qty{0.5}{} & & calculated \\ \hline \hline
         $\Psi$ & \qty{0.86(25)}{} & \qty{25}{\%} & relative difference between theory and experiment
    \end{tabular}
    \caption{Parameters used to quantify the amplitude of the simulated correlation signal.
    All values contribute as factors in the correlation amplitude. 
    By performing Gaussian propagation of uncertainty we obtain a systematic amplitude uncertainty of \qty{25}{\%}, which is the uncertainty of the systematic deviation between our simulation data and the measurements $\Psi=0.86$.}
    \label{tab:parameters}
\end{table*}

To fit \cref{eq:corrampl_final} to the experiment, we have to account for two additional effects.
The $z$ profile of the modes and the sub-Nyquist sampling of the magnons due to the repetition rate of the laser system, which lead to additional correction factors $\psi_f$ and $\psi_\mathrm{p}$ in the integrand of \cref{eq:corrampl_final}.

\paragraph{Mode Profiles}
So far, we have assumed a single mode per in-plane $\vec{k}$, which does not account for $z$ modulation.
As our sample is $\mathrm{\mu m}$ thick, we assume those higher order modes have approximately the same frequency but a different $z$ profile.
The $z$ profile of all modes $j$ is $p_{j}(z) = \sin\left( \frac{j\pi}{d^{\mathrm{z}}} \left( z+\frac{d^{\mathrm{z}}}{2} \right) \right)$, assuming that the full amplitude in the middle of the sample determines the magnon number.

The integral over the sample is then given by
\begin{align}
    \frac{1}{d^{\mathrm{z}}} \int_{-\frac{d^{\mathrm{z}}}{2}}^{\frac{d^{\mathrm{z}}}{2}} p_{j}(z)\mathrm{d}z = \frac{2}{j\pi} \text{  } (j \text{ even}) \text{ or } 0 \text{  } (j \text{ uneven}).
\end{align}
Due to the correlation, this term enters the final signal squared.
Thus, for each $\vec{k}$ we determine the additional factor $\psi_{\mathrm{p}}$ by summing over all modes $j$ as
\begin{align}
    \psi_{\mathrm{p}} = \frac{4}{\pi^2} \sum_{i=0}^\infty \frac{1}{(2i+1)^2} =\frac{1}{2}.
\end{align}
So, for FemNoC, we are mainly sensitive to the $j=1$ mode, while higher-order modes only contribute with a small correction factor due to the integration over their alternating $z$ profile.

\paragraph{Sub-Nyquist sampling with the repetition rate}
In our sample all magnons are long-lived compared to the repetition rate of the laser system.
Therefore, we have to account for the fact that FemNoC only samples the magnon population at discrete time points with a spacing of $\delta t = \frac{1}{f_{\mathrm{rep}}}$, where $f_{\mathrm{rep}}$ is the repetition rate of the laser system.
This leads to modes with a frequency $f_{\vec{k}}$ close to integer multiples of $f_{\mathrm{rep}}$ being sampled without producing pulse-to-pulse variations and subsequently not contributing to the correlation function.
Thus, only a part of the available frequencies contribute to the correlation function, resulting in a correction factor of $\psi_{f} = \qty{0.127(5)}{}$ to account for this effect.
We determine this value by measuring the coherent response of the sample to a microwave drive at different frequencies (see SM for details).

\paragraph{Ellipticity}

For our calculation, we have assumed circular precession of the magnetization as a first-order approximation for the dynamics of the magnon modes.
In general, the precession is elliptic, which leads to a reduced out-of-plane component of the magnetization and a reduced correlation amplitude respectively.

A full treatment of the ellipticity of the precession we leave for future work, as it requires a more detailed analysis of the magnon modes in the sample (e.g. by using micromagnetic simulations) to calculate the stiffness fields of the modes at different k-vectors.

As an approximation for the effects of ellipticity, we can assume a global ellipticity factor $\epsilon = \frac{M^z}{M^x}$.
Reformulating \cref{eq:circ_ampl_estimation} accordingly to 
\begin{align}
    \langle M^z(t)^2 + M^x(t)^2 \rangle_t = M_{\mathrm{sat}}^2 - \left(M_{\mathrm{sat}} - \frac{\hbar \gamma}{V} n_{\vec{k}}\right)^2
\end{align}
results in
\begin{align}
    a_{\vec{k}}^2 \approx \frac{2 M_{sat} \hbar \gamma \epsilon}{V} n_{\vec{k}} \label{eq:elliptic_ampl_estimation}.
\end{align}
The final correlation function is thus
\begin{align}
    C(\Delta t) = \frac{M_\mathrm{sat} \hbar \gamma \epsilon}{4 \pi^2 d^{\mathrm{z}}} \int \mathrm{d}^2 \vec{k} \, n_{\vec{k}} \cos(2\pi f_{\vec{k}} \Delta t) s_{\vec k}  \psi_{f} \psi_p \Psi. \label{eq:corrampl_final_elliptic}
\end{align}

The value of $\epsilon$ for the Kittel-mode can be calculated as 
\begin{align}
    \epsilon = \sqrt{\frac{H}{H+M_{\mathrm{eff}}}}
\end{align} directly following from \cite{kittelTheoryFerromagneticResonance1948}.
Using the effective magnetization $M_{\mathrm{eff}} = \qty{60.9(5)}{\kilo\A\per\m}$, we would obtain a maximum ellipticity of $\epsilon = 0.78$ at the lowest field before the emergence of domains $B_{\mathrm{ext}} = \qty{120}{\milli\tesla}$, and a minimal ellipticity of $\epsilon = 0.91$ at the highest field $B_{\mathrm{ext}} = \qty{350}{\milli\tesla}$.

This value fits well to the remaining systematic deviation $\Psi = 0.86(25)$, we observe between the simulated and measured correlation amplitude.

\paragraph{Remaining Deviation} In total, a systematic deviation factor of $\Psi = \qty{0.86(25)}{}$ remains between the simulated results and the experimental data, which we included in all presented fits.
We attribute this deviation to expected errors in the calibration procedure that convert our measured data to units of magnetization and the factors in \cref{eq:corrampl_final}.
For the conversion, these are the parameters $C_{\alpha, \mathrm{ave}}$ and $C_M$, which are described in detail in Ref. \cite{weissQuantifyingAmplitudesUltrafast2025}.

All of these values (listed in \cref{tab:parameters}) are multiplied together to result in the amplitude of the simulation and data, leaving room for a systematic deviation of $\pm \qty{25}{\%}$ affecting only the amplitude and not the shape of the signal.
Thus, our resulting $\Psi$ can be reconciled with the value 1 within the scope of uncertainties -- in other words, our model quantitatively agrees with the measured data.
Beyond this quantified uncertainty, we expect a systematic tendency of the measured signal to be smaller than the simulated one due to the ignored effects of ellipticity (discussed above).
Furthermore, we assume a perfect Gaussian beam profile and perfect spatial overlap of the beams in the sample.
Experimentally caused deviations from this ideal case would lead to a reduced correlation of the probes and thus a smaller signal than the idealized model.

The main contributors to the systematic uncertainty of $\Psi$ are the uncertainty of the sample thickness, the calibration of the Verdet constant, and the calibration of the electronic detection scheme (discussed in \cite{weissQuantifyingAmplitudesUltrafast2025}).
The electronic calibration, as well as the value for $\psi_{\mathrm{f}}$ do not depend on the sample and are thus transferable.
The Verdet constant, saturation magnetization, sample thickness and mode profile $\psi_{\mathrm{p}}$ are sample specific, and thus have to be determined for each sample individually.
In total their combined error would comprise a different exact value of $\Psi$ close to 1.
\end{document}


\preprint{Version 2026.09.07}

\renewcommand{\d}{\ensuremath{\mathrm{d}}}

\newcommand{\stbg}[1]{{\color[rgb]{1,0,0} #1}}
\newcommand{\rs}[1]{{\color[rgb]{0,0.7,0.7} #1}}
\newcommand{\fh}[1]{{\color[rgb]{0,0,1} #1}}
\newcommand{\move}[1]{{\color[rgb]{0,0,1} #1}}
\newcommand{\remove}[1]{{\color[rgb]{0,0.5,0} #1}}

\title{\textbf{Quantifying thermal and driven magnon populations with femtosecond noise correlation spectroscopy}}

\author{F.~S.~Herbst}
    \affiliation{Department of Physics, University of Konstanz, 78457 Konstanz, Germany}
\author{M.~A.~Weiss}
    \affiliation{Department of Physics, University of Konstanz, 78457 Konstanz, Germany}
\author{A.~Leitenstorfer}
    \affiliation{Department of Physics, University of Konstanz, 78457 Konstanz, Germany}
\author{M.~Lammel}
    \affiliation{Department of Physics, University of Konstanz, 78457 Konstanz, Germany}
\author{N.~Beaulieu}
    \affiliation{LabSTICC, CNRS, Université de Bretagne Occidentale, Brest 29238, France}
\author{J.~Ben~Youssef}
    \affiliation{LabSTICC, CNRS, Université de Bretagne Occidentale, Brest 29238, France}
\author{R.~Schlitz}
    \affiliation{Department of Physics, University of Konstanz, 78457 Konstanz, Germany}
\author{S.~T.~B.~Goennenwein}
    \affiliation{Department of Physics, University of Konstanz, 78457 Konstanz, Germany}

\date{07. September 2026}

\onecolumngrid
\section{Supplementary Materials}

\subsection{Faraday microscopy data} \label{subsec:faraday}

\begin{figure*}[h]
    \centering
    \includegraphics[width=\textwidth]{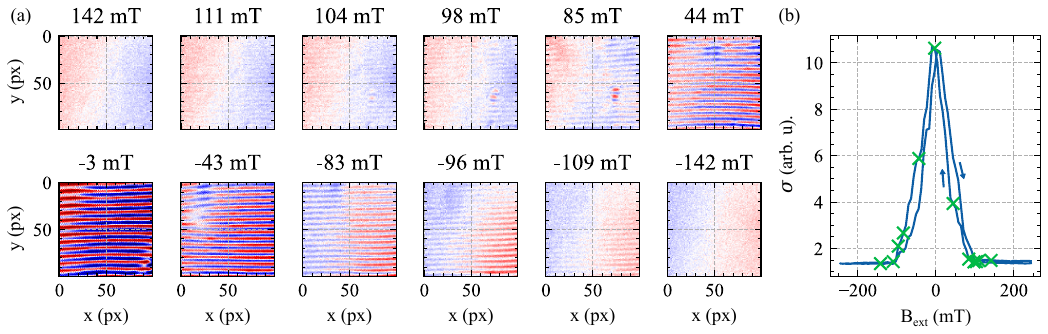}
    \caption{
        (a) Faraday microscopy images of the Bi:YIG sample as a function of in-plane magnetic field $B_{\mathrm{ext}}$.
        The colors are normalized to the median value of all images to be zero.
        (b) Standard deviation $\sigma$ of the pixel intensity values as a function of the in-plane magnetic field.
        Due to the emergence of domains, $\sigma$ increases significantly at fields below $\approx \SI{120}{\milli\tesla}$.
    }
    \label{fig:faraday}
\end{figure*}

To quantify the domain properties of the Bi:YIG sample, we performed Faraday microscopy measurements as a function of the applied in-plane magnetic field.
We use a custom-built microscope setup integrated into the FemNoC system.
The incoherent, red light of an LED ($\lambda = \qty{629}{\nano\meter}$) is linearly polarized and collimated to illuminate the sample.
The transmitted light is then collected by the microscope objective and detected by a CMOS camera.
Faraday contrast is obtained by using a polarizer in front of the camera, which is slightly detuned from the crossed position to the incident polarization.
The resulting images after subtracting a background image without domain contrast are shown in \cref{fig:faraday} (a) as a function of the applied in-plane magnetic field.

We find a clear signature of stripe domains at low fields, which decrease in amplitude with higher fields but are still present up to fields of approximately \SI{120}{\milli\tesla}.
To achieve a more precise estimate of the transition point, we calculate the standard deviation of the pixel intensity values $\sigma$ in each image as shown in \cref{fig:faraday} (b).
The domains lead to a significant increase of $\sigma$ at low fields, which vanishes around \SI{120}{\milli\tesla}, quantitatively determining the transition point between the single-domain and multi-domain states of the sample.

\subsection{Kittel Fit}

We fit the observed resonance frequencies at high fields with the Kittel formula for in-plane magnetized thin films \cite{kittelTheoryFerromagneticResonance1948}, which is given by
\begin{align}
    f = \frac{\gamma}{2\pi} \sqrt{B_{\mathrm{ext}} \cdot (B_{\mathrm{ext}} + \mu_0 M_{\mathrm{eff}})}
\end{align}
where $B_{\mathrm{ext}}$ is the external magnetic field, $M_{\mathrm{eff}}$ is the effective magnetization, and $\gamma$ is the gyromagnetic ratio.
We find $\gamma= \qty{28.1(1)}{\giga\hertz\per\tesla}$ and $M_{\mathrm{eff}} = \SI{60.9(5)}{\kilo\ampere\per\m}$ (black dashed line in Fig. 3 of the main text).

\subsection{Effects of varying probe power}

\begin{figure*}
    \centering
    \includegraphics[width=\textwidth]{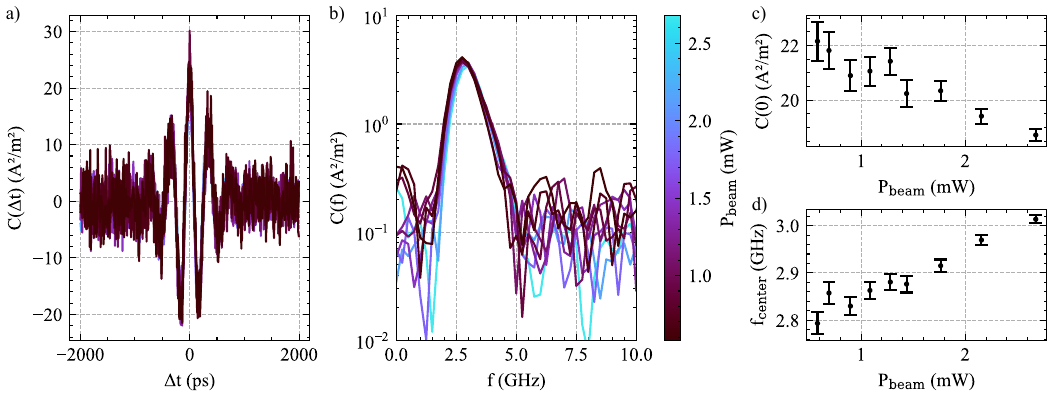}
    \caption{
        (a) $C(\Delta t)$ of the thermal background for different probe powers
        (b) Fourier transform $C(f)$ of the correlation function for different probe powers.
        (c) Extracted values for $C(0)$ and the central magnon frequency $f_{\mathrm{center}}$ as a function of probe power.
    }
    \label{fig:beampowerdependence}
\end{figure*}

We check the effects of varying the probe power on the measured correlation function $C(\Delta t)$ by using a variable neutral density filter ahead of the sample.
We perform measurements at different probe powers $P_{\mathrm{beam}}$, which is the total power of both detection branches measured at the photodiodes.
The results are shown in \cref{fig:beampowerdependence} where the $P_{\mathrm{beam}}$ is varied from \SI{0.5}{\milli\watt} to \SI{3}{\milli\watt}.

The signal-to-noise ratio (SNR) of the measurement scales linearly with the probe power \cite{weissQuantifyingAmplitudesUltrafast2025}.
To achieve roughly the same SNR in all measurements, the data at low probe powers is averaged longer ($T_{\mathrm{avg}} \propto 1/P_{\mathrm{beam}}^2$).

We find that the sample absorbs about \SI{9}{\percent} of the laser probe power (measured before the focusing optics), while additional losses of $\approx \SI{50}{\percent}$ can be attributed to clipping at the CPW, reflections at the optical components, and reflections at the sample surfaces.
The remaining \SI{41}{\percent} are transmitted through the detection branches and detected by the photodiodes giving the measured probe power $P_{\mathrm{beam}}$.

For our measurements, which were performed at $P_{\mathrm{beam}} = \SI{1.5}{\milli\watt}$, we thus estimate that the sample absorbs an average power of \SI{0.3}{\milli\watt}.

The observed magnetization correlation amplitude show continuous increase of the central magnon frequency $f_{\mathrm{center}}$ with increasing probe power, and a subsequent decrease of the correlation amplitude $C(0)$.
As this effect emerges with changing probe power, we attribute it to local heating of the sample by the laser probe.
In the entire range of probe powers, the shifts are in the order of \SI{10}{\percent} of the absolute values.
We thus conclude that while there is a small influence of local heating, the results shown in the main text are not significantly altered from the true thermal magnon background.

\subsection{Determination of the frequency correction factor}

As we discuss in the appendix, the measured correlation amplitude $C(0)$ is reduced due to the finite frequency resolution of the measurement.
This originates from the fact that the FemNoC measurement scheme is only sensitive to pulse-to-pulse fluctuations of the magnetization.
Modes with frequencies at multiples of the laser repetition rate $f_{\mathrm{rep}}$ remain undetected, as they have the same phase at each laser pulse and thus do not produce any pulse-to-pulse fluctuations.
This leads to a reduction of the measured correlation amplitude $C(0)$ compared to the true correlation amplitude, which we correct for by introducing a frequency correction factor $\psi_f$.

To determine $\psi_f$ we excite the sample with a coherent microwave signal.
We sweep the excitation frequency around \SI{5.82}{\giga\hertz} and \SI{5.9}{\giga\hertz}, which are close to the resonance frequency.
The microwave leads to a precession of the magnetization at the excitation frequency $f_\mathrm{dr}$.

\begin{figure}[h]
    \centering
    \includegraphics[width=0.4\textwidth]{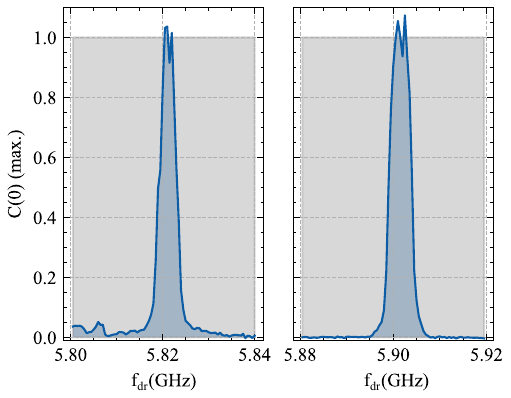}
    \caption{Example traces of the measured correlation amplitude $C(0)$ as a function of driving frequency $f_\mathrm{dr}$ (blue), normalized to the maximum value. On average the area under the curve is only a fraction of $\psi_f = \qty{0.127(5)}{}$ of the full area, which is determined by the gray envelope, determining the frequency correction factor $\psi_f$.
    }
    \label{fig:determine_fcorrectionfactor}
\end{figure}

\Cref{fig:determine_fcorrectionfactor} shows the measured correlation amplitude $C(0)$ as a function of the excitation frequency.
Due to the pulse-to-pulse fluctuation measurement scheme of FemNoC, the setup is only sensitive to the magnetization dynamics at frequencies that are odd integer multiples of half the laser repetition rate $f_{\mathrm{rep}} = f_{\mathrm{rep}} \cdot (m + 0.5)$, where $m$ is an integer.
In \cref{fig:determine_fcorrectionfactor} this leads to peaks in the measured correlation amplitude at $f = \qty{5.82}{\giga\hertz}$, and \qty{5.9}{\giga\hertz}.
We now extract the average correlation amplitude $C(0)$ at each peak and draw a window (gray) of width $\Delta f = \qty{40}{\mega\hertz}$ around the peak, which approximates the theoretical amplitude if the measurement was performed with infinite frequency resolution.
We then calculate the ratio of the measured correlation amplitude $C(0)$ to the average theoretical amplitude in the window, which gives us the correction factor $\psi_f = 0.127(5)$.

Although we show only a limited frequency range in \cref{fig:determine_fcorrectionfactor}, the correction factor $\psi_f$ is independent of frequency, as it is only determined by the laser repetition rate and the applied demodulation settings.

\bibliography{main}